\documentclass[footinbib,aps,twocolumn,superscriptaddress,floatfix,prl]{revtex4-2}

\usepackage[utf8]{inputenc}
\usepackage{amssymb,amsmath,amsfonts,accents,amsthm}
\usepackage{mathtools}
\usepackage{braket}
\usepackage[colorlinks=true,citecolor=blue,linkcolor=blue,urlcolor=blue]{hyperref}
\usepackage{xcolor}

\usepackage[german,english]{babel}

\makeatletter
\def\bbl@set@language#1{%
	\edef\languagename{%
		\ifnum\escapechar=\expandafter`\string#1\@empty
		\else\string#1\@empty\fi}%
	\@ifundefined{babel@language@alias@\languagename}{}{%
		\edef\languagename{\@nameuse{babel@language@alias@\languagename}}%
	}%
	\select@language{\languagename}%
	\expandafter\ifx\csname date\languagename\endcsname\relax\else
	\if@filesw
	\protected@write\@auxout{}{\string\select@language{\languagename}}%
	\bbl@for\bbl@tempa\BabelContentsFiles{%
		\addtocontents{\bbl@tempa}{\xstring\select@language{\languagename}}}%
	\bbl@usehooks{write}{}%
	\fi
	\fi}
\newcommand{\DeclareLanguageAlias}[2]{%
	\global\@namedef{babel@language@alias@#1}{#2}%
}
\makeatother

\DeclareLanguageAlias{en}{english}
\DeclareLanguageAlias{English}{english}
\DeclareLanguageAlias{Englisch}{english}
\DeclareLanguageAlias{EN}{english}
\DeclareLanguageAlias{en-US}{english}
\DeclareLanguageAlias{de}{german}

\usepackage{graphicx}
\usepackage[export]{adjustbox}

\usepackage{xcolor}
\newcommand{\PRLSec}[1]{\textit{#1}.---}

\usepackage[most]{tcolorbox}
\tcbuselibrary{theorems}
\usepackage{cleveref}
\newtcbtheorem[	crefname={theorem}{Theorems},Crefname={theorem}{Theorem}]{Theorem}{Theorem}{
	colback=green!5,
	colframe=green!35!black,
	fonttitle=\bfseries,
	breakable,
}{thm}
\newtcbtheorem[	crefname={lemma}{Lemma},Crefname={lemma}{Lemma}]{Lemma}{Lemma}{
	colback=orange!5,
	colframe=orange!35!black,
	fonttitle=\bfseries,
	breakable,
}{lem}

\newtcbtheorem[	crefname={corollary}{Corollary},Crefname={corollary}{Corollary}]{Corollary}{Corollary}{
	colback=orange!5,
	colframe=orange!35!black,
	fonttitle=\bfseries,
	breakable,
}{cor}

\newtcbtheorem[
crefname={definition}{Definition},
Crefname={definition}{Definition}]
{Definition}{Definition}{
	colback=blue!5,
	colframe=blue!35!black,
	fonttitle=\bfseries,
}{def}

\newcommand{\sbb}{\sbar{}\sbar}
\newcommand{\ssb}{S\sbar}
\newcommand{\sbs}{\sbar{}S}

\newcommand{\sbar}{\overline{S}}

\newcommand{\ISR}[2]{\mathcal{R}_{#1}(#2,\lambda)}

\newcommand{\ham}{H}

\newcommand{\qMat}[2]{Q^{(#1,#2)}}

\newcommand{\comm}[2]{\left[#1,#2\right]}

\newcommand{\smRef}{See Supplemental Material for mathematical details and proofs, which includes also Refs. \cite{Morfonios2020aCospectralityPreservingGraphModifications,Chan2020AQFundamentalsFractionalRevivalGraphs,Eisenberg2019DM3422821PrettyGoodQuantumState,Tee2005RLIMS8123EigenvectorsBlockCirculantAlternating,Vanderbilt2018BerryPhasesElectronicStructure}.}

\begin{document}
	
	\title{Latent symmetry induced degeneracies}
	
	\author{M. Röntgen}
	\affiliation{%
		Zentrum für optische Quantentechnologien, Universität Hamburg, Luruper Chaussee 149, 22761 Hamburg, Germany
	}%
	
	\author{M. Pyzh}%
	\affiliation{%
		Zentrum für optische Quantentechnologien, Universität Hamburg, Luruper Chaussee 149, 22761 Hamburg, Germany
	}%
	
	\author{C. V. Morfonios}%
	\affiliation{%
		Zentrum für optische Quantentechnologien, Universität Hamburg, Luruper Chaussee 149, 22761 Hamburg, Germany
	}%
	
	\author{N. E. Palaiodimopoulos}%
	\affiliation{%
		Department of Physics, University of Athens, 15771 Athens, Greece
	}%
	\author{F. K. Diakonos}%
	\affiliation{%
		Department of Physics, University of Athens, 15771 Athens, Greece
	}%
	
	\author{P. Schmelcher}
	\affiliation{%
		Zentrum für optische Quantentechnologien, Universität Hamburg, Luruper Chaussee 149, 22761 Hamburg, Germany
	}%
	\affiliation{%
		The Hamburg Centre for Ultrafast Imaging, Universität Hamburg, Luruper Chaussee 149, 22761 Hamburg, Germany
	}%

	\begin{abstract}
		Degeneracies in the energy spectra of physical systems are commonly considered to be either of accidental character or induced by symmetries of the Hamiltonian.
		We develop an approach to explain degeneracies by tracing them back to symmetries of an isospectral effective Hamiltonian derived by subsystem partitioning.
		We provide an intuitive interpretation of such latent symmetries by relating them to corresponding local symmetries in the powers of the underlying Hamiltonian matrix.
		As an application, we relate the degeneracies induced by the rotation symmetry of a real Hamiltonian to a non-abelian latent symmetry group.
		It is demonstrated that the rotational symmetries can be broken in a controlled manner while maintaining the underlying more fundamental latent symmetry.
		This opens up the perspective of investigating accidental degeneracies in terms of latent symmetries.
	\end{abstract}
	
	\maketitle
	
	\PRLSec{Introduction}Identifying the origin of spectral degeneracies in quantum systems is of fundamental importance for the understanding and control of their structural and dynamical properties.
	Degenerate states are at the heart of spectacular phenomena like the Jahn-Teller effect \cite{Bersuker2010JahnTellerEffect} and the quantum Hall effect \cite{Ezawa2013QuantumHallEffectsRecentTheoretical,vonKlitzing2017ARCMP813QuantumHallEffectDiscovery} as well as the electromagnetic response of e.g., atoms or molecules \cite{Phillips1998RMP70721NobelLectureLaserCooling,Baer2002124RoleDegenerateStatesChemistry} in general.
	In lattice systems designed macroscopic degeneracies can realize flat bands within a variety of setups including optical lattices, photonic waveguide arrays, and  superconducting networks \cite{Leykam2018AP31473052ArtificialFlatBandSystems}. 
	Further, degeneracies in the form of conical intersections of molecular potential energy surfaces play a central role for ultrafast dynamical decay processes \cite{Domcke200415ConicalIntersectionsElectronicStructure,Koppel1984AiCP5759MultimodeMolecularDynamicsBornOppenheimer} and are responsible e.g., for molecular self-repair mechanisms in photobiology \cite{Domcke2012ARPC63325RoleConicalIntersectionsMolecular}. 
	
	When degeneracies occur in the energy spectrum, the first place to seek their origin is commonly the group of geometrical symmetry operations commuting with the underlying Hamiltonian.
	Prominent examples for such symmetries are the molecular point group in chemistry or the space group in crystallography.
	If this group is non-abelian---that is, if at least two symmetry operations do not commute with each other---it induces degeneracies of multiplicities determined by the dimensions of the group's irreducible representations.
	More challenging is the reverse question of assigning degeneracies to a symmetry group with a physical significance
	\cite{McIntosh1959AJP27620AccidentalDegeneracyClassicalQuantum,McIntosh1971GTaiA2SymmetryDegeneracy}.
	A famous example of a physically significant, yet not obvious, symmetry from the early days of quantum theory is the $SO(4)$ symmetry leading to the conservation of the Runge-Lenz vector in the Hydrogen atom\cite{Fock1935ZP98145ZurTheorieWasserstoffatoms}.
	If no such physically meaningful symmetry group can be found, the degeneracy is traditionally called accidental \cite{vonNeumann1929PZ30467UberVerhaltenEigenwertenBei}.
	This often occurs for systems with several or many degrees of freedom where eigenenergies happen to coincide at some location in the corresponding parameter space, intersections of molecular potential energy surfaces being a typical example \cite{Yarkony1996RMP68985DiabolicalConicalIntersections}.
	
	In this work, we promote a different viewpoint on assigning degeneracies to symmetries of the system.
	Instead of performing a symmetry analysis of the Hamiltonian itself, we do this for the effective Hamiltonian obtained from the original one by reducing it onto a subsystem while retaining the energy spectrum.
	We note that its core property---the preservation of the energy spectrum---clearly distinguishes this approach from those which analyze the symmetries of an effective model obtained by truncation or a mean-field ansatz.
	Focusing on generic discrete models, we here show how geometrical symmetries of the isospectrally reduced Hamiltonian induce spectral degeneracies for the original system.
	Such latent symmetries, as introduced recently in graph theory \cite{Smith2019PA514855HiddenSymmetriesRealTheoretical}, are generally not apparent in the original system at hand.
	In fact, as we show here, they are directly linked to corresponding local symmetries, though in all powers of the original Hamiltonian.
	Navigating through the proposed concepts, visualized by minimalistic examples, we (i) show how non-abelian latent symmetries are necessarily induced by rotation symmetries of a real Hamiltonian, and (ii) demonstrate that these latent symmetries, along with their induced degeneracies, can be preserved even when breaking the original rotational symmetry. 
	Lastly, we link a special case of latent symmetry to what we call here a generalized exchange symmetry of the Hamiltonian.
	
	\begin{figure}[htb] 
		\centering
		\includegraphics[max size={\columnwidth}]{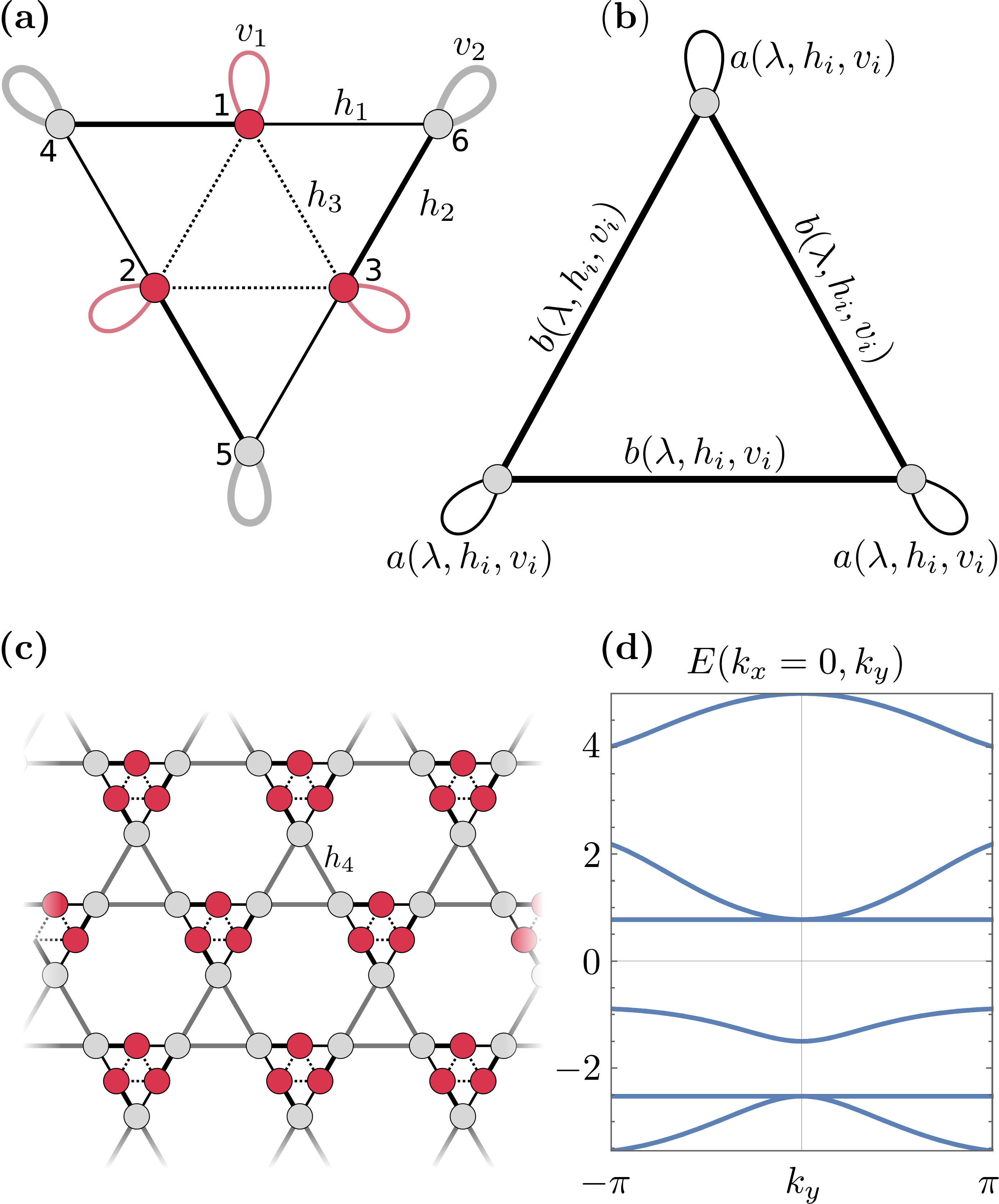}
		\caption{\textbf{(a)} A six-site Hamiltonian $\ham$ which features a non-abelian $D_{3}$ permutation symmetry if $h_{1} = h_{2}$, but only an abelian $C_{3}$ permutation symmetry if $h_{1} \ne h_{2}$. A line between two different sites $i,j$ corresponds to a non-vanishing matrix element $\ham_{i,j}$, taking parametric values $h_{1},h_{2}$, or $h_{3}$ (indicated by different line styles). Loops connecting a site to itself correspond to diagonal matrix elements $H_{i,i}$ with parametric values $v_{1}$ or $v_{2}$.
			\textbf{(b)} The result of the isospectral reduction of $\ham$ over the three red sites $S = \{1,2,3\}$.
			The reduced Hamiltonian [\cref{eq:ISRintroductoryExample}] features a $D_{3}$ permutation symmetry for any choice of $\lambda, h_{i}$, or $v_{i}$.
			(\textbf{c}) A modified Kagome-lattice with $\ham$ as a unit cell.
			The band structure of this lattice for $k_{x} = 0$ is plotted in (\textbf{d}) for $h_{1} = 4/3$, $h_{2} = 5/3$, $h_{3} = 0.7$, $h_{4} = 3/2$, $v_{i} = 0$.
		}
		\label{fig:introductoryExample}
	\end{figure}
	
	\PRLSec{Degeneracies from latent symmetries}%
	The concepts and results developed in this work are valid for generic setups described by a finite-dimensional matrix.
	This matrix can be drawn from a wide range of physical platforms: It could represent a Bloch Hamiltonian of a tight-binding lattice \cite{CrastodeLima2020PRB101041107HighdegeneracyPointsProtectedSitepermutation}, a molecular Hückel Hamiltonian \cite{Herndon1975T3199IsospectralGraphsMolecules,Levine2000QuantumChemistry}, a multiport scattering matrix \cite{Richoux2020JPDAP53235101MultifunctionalResonantAcousticWave}, or very generally the matrix $\ham$ occurring in (linearized) dynamical problems \cite{Galor2007DiscreteDynamicalSystems}, such as coupled oscillators \cite{Wilson1980MolecularVibrationsTheoryInfrared}.
	To convey the main ideas in a transparent way, we will illustrate it by means of minimalistic prototypical setups.
	
	In order to reveal the latent symmetries of a general complex matrix $\ham$, we will rely on a dimensional reduction of $\ham$ which preserves the eigenvalue spectrum.
	This \emph{isospectral} reduction is defined as \cite{Smith2019PA514855HiddenSymmetriesRealTheoretical,Bunimovich2014IsospectralTransformationsNewApproach}
	\begin{equation} \label{eq:ISRDefinition}
	\ISR{S}{\ham} = \ham_{SS} - \ham_{\ssb} \left(H_{\sbb} - \lambda I \right)^{-1} \ham_{\sbs} ,
	\end{equation}
	whereby $S$ is a set sites and $\sbar$ denotes the complement set of all other sites of the given  setup.
	$\ham_{SS}$ and $\ham_{\sbb}$ denote the respective Hamiltonians of the sub-systems consisting only of the sites in $S$ or $\sbar$.
	$\ham_{\sbs}$ and $\ham_{\ssb}$ represent the coupling between the two sub-systems, and $I$ is the identity matrix.
	The isospectral reduction $\ISR{S}{\ham}$ is equivalent to an effective Hamiltonian gained from a subsystem partitioning of $\ham$ \cite{Grosso2013SolidStatePhysics}, and its entries are rational functions of the parameter $\lambda$.
	
	A Hamiltonian $\ham$ is \emph{latently symmetric} if there exists an isospectral reduction $\ISR{S}{\ham}$ with a symmetry, that is, which commutes with a group of matrices $\{M\}$ independent of $\lambda$.
	We now demonstrate this concept by means of the simple $6$-site Hamiltonian $\ham$ depicted in \cref{fig:introductoryExample} (a).
	This Hamiltonian illustrates the minimal prototype of a system with non-trivial latent symmetry.
	$\ham$ is parametrized by three real coupling parameters $h_{i} \ne 0$, $i \in \{1,2,3\}$ and two on-site potentials $v_{1},v_{2}$.
	The eigenvalue spectrum of $\ham$ contains two doubly degenerate eigenvalues for any choice of these parameters.
	To explain these degeneracies in terms of latent symmetries of $\ham$, we reduce it by means of \cref{eq:ISRDefinition} over $S = \{1,2,3\}$.
	This yields the symmetric matrix
	\begin{equation} \label{eq:ISRintroductoryExample}
	\ISR{S=\{1,2,3\}}{\ham} = \begin{pmatrix}
	a & b & b \\
	b & a & b \\
	b & b & a
	\end{pmatrix},
	\end{equation}
	with $a = v_{1} + \frac{h_1^2+h_2^2}{\lambda - v_{2}}$, $b = \frac{h_1 h_2}{\lambda -v_{2}}+h_3$.
	A graphical representation of \cref{eq:ISRintroductoryExample} is depicted in \cref{fig:introductoryExample} \textbf{(b)}.
	The graph is highly symmetric and is invariant under six symmetry operations: three rotations and three reflections.
	These six operations form the so-called dihedral group $D_{3}$, which is non-abelian.
	
	We now draw a general connection between non-abelian latent symmetries of a given Hamiltonian $\ham$ and its eigenvalue spectrum.
	To this end, we use the fact that \emph{each} of the so-called ``nonlinear'' eigenvalues belonging to $\ISR{S}{\ham}$ in \cref{eq:ISRintroductoryExample}, defined as the solutions $\lambda_{j}$ to the nonlinear eigenvalue problem
	\begin{equation}
	Det\left(\mathcal{R}_{S}(\ham,\lambda_{j}) - \lambda_{j} I \right) = 0
	\end{equation}
	is also an eigenvalue of $\ham$ \cite{Bunimovich2014IsospectralTransformationsNewApproach}.
	Moreover, whenever the eigenvalue spectra of $\ham$ and of the subsystem $\ham_{\sbb}$ do not intersect, the eigenvalue spectra of $\ISR{S}{\ham}$ and $\ham$ coincide \cite{Bunimovich2014IsospectralTransformationsNewApproach}.
	This motivates calling $\ISR{S}{\ham}$ an ``isospectral reduction''.
	From the above considerations, it is clear that degeneracies in the eigenvalue spectrum of $\ISR{S}{\ham}$ necessarily correspond to degeneracies in the eigenvalue spectrum of $\ham$.
	Moreover, and as we show in Sec. I. of the Supplemental Material of this work \cite{Note1}, non-abelian symmetries of the isospectral reduction $\ISR{S}{\ham}$ lead to degeneracies in the spectrum of its nonlinear eigenvalues.
	\emph{Thus, non-abelian latent symmetries of $\ham$ necessarily induce degeneracies onto the eigenvalue spectrum of $\ham$.}
	Specifically, lower bounds on the multiplicity of $\ham$'s eigenvalues are given by dimensions of the irreducible representations of the underlying non-abelian symmetry group of $\ISR{S}{\ham}$.
	
	We emphasize that the above statements are completely general in the sense that they are valid for all kinds of latent symmetries (not just permutations), and for arbitrary (even non-hermitian) diagonalizable matrices $\ham$.
	Irrespective of this applicability to general symmetries, we concentrate on the special case of permutation symmetries throughout this Letter.
	After all, permutation symmetries are among the easiest to detect---often by bare eye---and thus provide a convenient workhorse for depicting the main features of latent symmetries.
	
	In the above, we have explained the spectral degeneracies of the prototype example \cref{fig:introductoryExample} (a) in terms of its latent symmetries.
	This system has been deliberately designed to be as simple as possible in order to convey the main ideas of latent symmetries.
	The underlying concept is, however, not limited to such basic examples, but can be applied to larger systems, as we demonstrate now.
	\Cref{fig:introductoryExample} (c) shows a lattice built by taking the prototype Hamiltonian $\ham$ of \cref{fig:introductoryExample} (a) as a unit cell.
	The band structure of this lattice is depicted in \cref{fig:introductoryExample} \textbf{(d)}.
	At the $\Gamma$-point, that is, at $\mathbf{k} = 0$, the corresponding Bloch-Hamiltonian features the same latent symmetries as $\ham$ in \cref{fig:introductoryExample} (a).
	This explains the two double degeneracies in the band structure\footnote{\smRef}.
	Interestingly, the lattice further hosts two flat bands, which in general can also be designed through latent symmetries \cite{Morfonios2021aFlatBandsLatentSymmetry}.
	
	\PRLSec{Latent $D_{n}$ permutation symmetries}%
	Let us now examine the symmetries of the prototype example of \cref{fig:introductoryExample} in more detail.
	This setup is invariant under permutations which cyclically permute sets of three sites, graphically represented by rotations of multiples of $\frac{2\pi}{3}$.
	These rotations form the abelian cyclic group of order $3$, denoted by $C_{3}$.
	As we have seen above, the setup also featured a latent $D_{3}$ permutation symmetry, and this is no coincidence.
	Indeed, as we show in the Supplemental Material, \emph{every} $C_{n}$-permutation symmetric real Hamiltonian $\ham$ features a latent $D_{n}$ permutation symmetry \cite{Note1}.
	As is well known, the dihedral group $D_{n}$ is non-abelian for $n \ge 3$, so that the underlying Hamiltonian automatically features degeneracies.
	This gives an alternative explanation to those degeneracies, which are classically understood in terms of the combination of the abelian group $C_{n\ge 3}$ and the real-valuedness of $\ham$ which corresponds to a time-reversal symmetry of $\ham$ \cite{Landau1981QuantumMechanicsNonRelativisticTheory}.	
	
	\PRLSec{Latent $D_{n}$ symmetries without any permutation symmetries}%
	\begin{figure}[tb]
		\centering
		\includegraphics[max size={\columnwidth}]{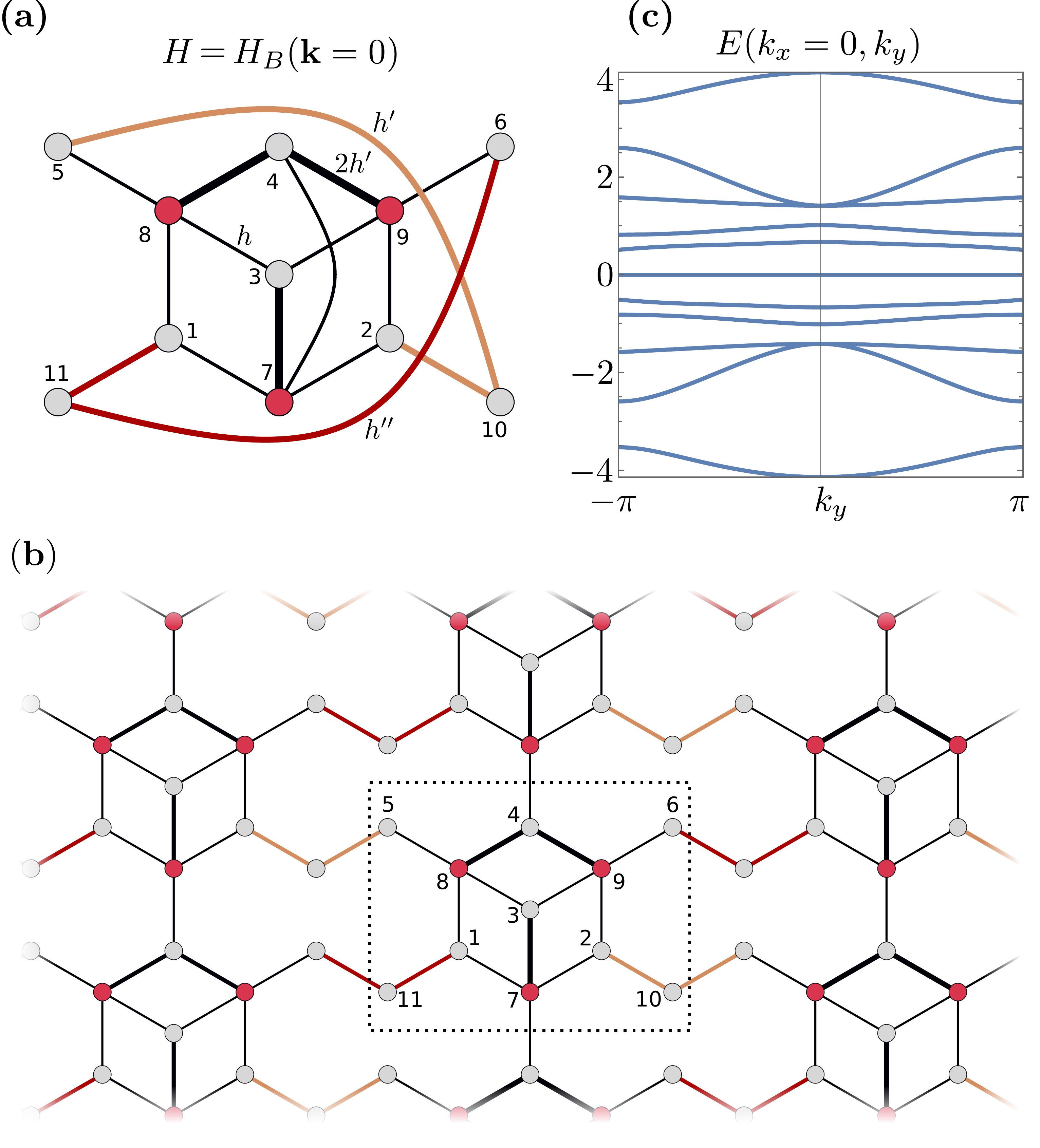}
		\caption{
			\textbf{(a)}	
			A Hamiltonian that features no permutation symmetry for $h\, h'\, h'' \ne 0$ and $h' \ne h''$.
			It does, however, feature a latent $D_{3}$ permutation symmetry that becomes visible when reducing over $S= \{7,8,9\}$.
			\textbf{(b)} A lattice whose Bloch-Hamiltonian $\ham_{B}$ at $\mathbf{k} = 0$ equals $\ham$. The dotted box shows the lattice unit cell.
			\textbf{(c)} The band structure of this lattice for $k_{x} = 0,h=1,h'=1/2,h''=3/4$.
		}
		\label{fig:noPermutations}
	\end{figure}
	Above we have stated that a $C_{n}$ permutation symmetry of a real Hamiltonian is a sufficient condition for a latent $D_{n}$ permutation symmetry.
	However, it is not a necessary condition.
	Indeed, we demonstrate in the following the versatility of latent symmetries by showing that they can even exist when the underlying Hamiltonian $\ham$ has \emph{no} permutation symmetry at all.
	\Cref{fig:noPermutations} (a) shows an example of such a Hamiltonian $\ham$, which can also be interpreted as the Bloch Hamiltonian $\ham_{B}(\mathbf{k} = 0)$ of the lattice in \cref{fig:noPermutations} (b) at crystal momentum $\mathbf{k} = 0$.
	A detailed derivation of this lattice is shown in Sec. V. of the Supplemental Material \cite{Note1}.
	For $h\, h' \, h'' \ne 0$ and $h' \ne h''$, $\ham$ does not feature any permutation symmetry.
	However, for any choice of those three hopping parameters, it features a latent $D_{3}$ symmetry which becomes visible when reducing $\ham$ over the three red sites $S=\{7,8,9\}$.
	As a result of this non-abelian latent symmetry, $\ham$ has at least one doubly degenerate eigenvalue pair for \emph{any} choice of $h'$ and $h''$.
	We can now understand the two double degeneracies in the band structure [depicted in \cref{fig:noPermutations} (c)] of the lattice of \cref{fig:noPermutations} (b) at $k_{x} = k_{y} = 0$: At this point, the Bloch-Hamiltonian is given by $\ham$, so that it features a latent $D_{3}$ symmetry and therefore also degeneracies.	
	
	Interestingly, when setting $h' = h''$, $\ham$ features a $C_{2}$ permutation symmetry, graphically corresponding to a reflection about the line connecting the sites $4$ and $7$.
	One can thus say that $h'$ and $h''$ are control parameters for a symmetry breaking, and since $\ham$ features a latent $D_{3}$ permutation symmetry for \emph{any} choice of $h,h',h''$, this opens the perspective of investigating and understanding symmetry breaking in terms of latent symmetries.
	In Sec. III. of the Supplemental Material, we show how latent symmetry preserving modifications (which may break permutation symmetries) can be derived \cite{Note1}.

	\PRLSec{Linking latent to local symmetries}%
	One might wonder if a latent symmetry leaves some recognizable traces in the original Hamiltonian.
	This is indeed the case: By expressing $\ISR{S}{\ham}$ as a power series in $\lambda$ and subsequently analyzing it order by order, one can show \cite{Note1} that 
	\begin{equation} \label{eq:equivalenceLatentLocalSymmetries}
	\comm{\ISR{S}{\ham}}{M} = 0 \;\;\Leftrightarrow \;\; \comm{\left(\ham^{k} \right)_{SS}}{M} = 0 \;\forall\;k
	\end{equation}
	where $M$ denotes a symmetry operation.
	In other words, symmetries of $\ISR{S}{\ham}$ correspond to local symmetries \cite{Kalozoumis2014PRL113050403InvariantsBrokenDiscreteSymmetries,Kalozoumis2013PRA87032113LocalSymmetriesOnedimensionalQuantum} of $\ham$ \emph{in all matrix powers}.
	In particular, $\ham$ itself has to be locally symmetric.
	Indeed, for our introductory example of \cref{fig:introductoryExample} \textbf{(a)} and $S = \{1,2,3\}$, we see that $\ham_{SS}$ denotes the inner triangle, which obviously features the same symmetries as the corresponxding isospectral reduction $\ISR{S}{\ham}$ depicted in \cref{fig:introductoryExample} \textbf{(b)}.
	
	\Cref{eq:equivalenceLatentLocalSymmetries} can be used to facilitate the search for latent permutation symmetries.
	To this end, let us assume that we are given a (possibly large) Hamiltonian $\ham$ and want to check if it features a latent permutation symmetry as the one depicted in \cref{fig:introductoryExample} (\textbf{b}).
	In other words, we look for a set of three sites $S = \{u,v,w\}$ such that $\ISR{S}{\ham}$ has the form of \cref{eq:ISRintroductoryExample}.
	Now, instead of computing and checking \emph{all possible} isospectral reductions of $\ham$ over three sites, we can use \cref{eq:equivalenceLatentLocalSymmetries} to see that any candidate sites $u,v,w$ necessarily have to fulfill $\left(\ham^{k} \right)_{u,u} = \left(\ham^{k} \right)_{v,v} = \left(\ham^{k} \right)_{w,w}$ for all $k$.
	This condition can be augmented by employing the Cayley-Hamilton theorem, which states that any matrix power $\ham^{k\ge N}$ ($N$ being the dimension of $\ham$) is a polynomial in smaller powers.
	Thus, by computing the matrix powers $\ham,\ham^{2},\ldots{},\ham^{N-1}$---the cost of which grows polynomially with $N$---and grouping the sites accordingly, the number of possible candidate sites $\{u,v,w\}$ can be drastically reduced.
	In particular, if there is any $k$ such that $\ham^{k}$ features no three sites with equal on-site potential $\left(\ham^{k}\right)_{i,i}$, a latent symmetry of the kind \cref{eq:ISRintroductoryExample} is impossible.
	
	\PRLSec{Generalized exchange symmetries}%
	Having demonstrated the relation of latent symmetries to symmetries of the subsystem $\ham_{SS}$ and to degeneracies of $\ham$, we finally relate a subclass of latent symmetries to symmetries of the original Hamiltonian $\ham$.
	This subclass consists of latent permutation symmetries of real Hamiltonians.
	Using graph-theoretical tools \cite{Kempton2020LAIA594226CharacterizingCospectralVerticesIsospectral,Godsil2017A1StronglyCospectralVertices}, such Hamiltonians can be shown to necessarily feature what we call here a \emph{generalized exchange symmetry} (GES).
	A GES is an orthogonal symmetric matrix $\qMat{i}{j}$ fulfilling $[\qMat{i}{j},\ham] = 0$ and $(\qMat{i}{j})^2 = I$ and which exchanges the two sites $i,j$ while acting on the remaining sites as an orthogonal transformation.
	In the special case when this transformation is a pure permutation, $\qMat{i}{j}$ becomes a normal exchange symmetry, i.e., it acts \emph{on each} site either as the identity or as an exchange operator.
	To provide an impression of the GESs, we explicitly computed---by solving the equations derived from its defining properties--- $Q^{(1,2)}$ for the Hamiltonian of \cref{fig:introductoryExample} \textbf{(a)}:
	\begin{equation} \label{eq:Q12}
	Q^{(1,2)} = \left(
	\begin{array}{cccccc}
	0 & 1 & 0 & 0 & 0 & 0 \\
	1 & 0 & 0 & 0 & 0 & 0 \\
	0 & 0 & 1 & 0 & 0 & 0 \\
	0 & 0 & 0 & \frac{h_1 h_2}{d}
	& 1-\frac{h_1^2}{d} &
	\frac{h_1
		\left(h_1-h_2\right)}{d} \\
	0 & 0 & 0 & 1-\frac{h_1^2}{d}
	& \frac{h_1
		\left(h_1-h_2\right)}{d} &
	\frac{h_1 h_2}{d} \\
	0 & 0 & 0 & \frac{h_1
		\left(h_1-h_2\right)}{d} &
	\frac{h_1 h_2}{d} &
	1-\frac{h_1^2}{d} \\
	\end{array}
	\right)
	\end{equation}
	with $d = h_1^2-h_2 h_1+h_2^2$.
	Note that for the case of $h_{1} = h_{2}$ the GES $Q^{(1,2)}$ becomes the ordinary exchange symmetry which permutes $(1,2)$, $(5,6)$, and leaves $3$ and $4$ invariant, and therefore describes the reflection about the line that connects sites $3$ and $4$ in \cref{fig:introductoryExample}.
	However, in the case where $h_{1} \ne h_{2}$, this pure permutation symmetry is broken, whereas the more abstract GES persists.
	We note that, while the GESs as an abstract symmetry class persists, the \emph{matrix entries} of $Q^{(1,2)}$ depend on $h_{1}$ and $h_{2}$.
	This is an important difference to the latent $D_{3}$ permutation symmetry of $\ham$, whose matrix representation is independent of the values of $h_{i}$.
	
	Finally, let us note that one can use the above insights to prove the existence of degeneracies for real latently $D_{n\ge 3}$ permutation symmetric Hamiltonians in yet another way \cite{Note1}.
	Such Hamiltonians feature more than one GES, and by explicitly constructing them it can be shown that at least two of them do not commute with each other.
	Since the Hamiltonian $\ham$ commutes with both of these GESs, it directly follows that $\ham$ has to have at least one degenerate eigenvalue.
	It remains an open task to classify GESs using group-theoretical tools.

	\PRLSec{Conclusions}%
	We have provided a theoretical framework which connects non-abelian latent symmetries of generic discrete models to their spectral degeneracies.
	For the important class of latent permutation symmetries, our results may allow for a geometrical explanation of apparently accidental degeneracies.
	Moreover, by identifying latent symmetries as local symmetries of all powers of the Hamiltonian, our results additionally suggest a convenient method for finding these latent symmetries.
	We further demonstrate that it is possible to break symmetries of an original Hamiltonian whilst preserving its latent symmetry.
	This may inspire techniques to modify---or probe---a given system asymmetrically without affecting its degeneracy.
	
	Our considerations apply quite generally to physical systems possessing a discrete representation in terms of a finite-dimensional matrix.
	This includes, among others, tight-binding models, molecular Hamiltonians in truncated orbital bases, and multiport scattering setups.
	We therefore envision the applicability of our results in a broad variety of setups, contributing to the better understanding, design, and control of spectral degeneracies beyond conventional symmetries.
	
	\begin{acknowledgments}
		\PRLSec{Acknowledgements}
		M.\,P. is thankful to the `Studienstiftung des deutschen Volkes' for financial support in the framework of a scholarship.
		The authors thank G.\,M. Koutentakis and J.\,Schirmer for valuable comments on the manuscript.
		
		The first three authors M.\,R., M.\,P, and C.\,V.\,M. contributed equally to this work.
	\end{acknowledgments}

\end{document}


\begin{center}
		\large \textbf{Supplemental Material}
	\end{center}

In this supplemental material, mathematical details of the proofs for the results presented in the main text as well as the Bloch-Hamiltonians for the lattices depicted in Figs. 1 and 2 of the main text are included.
It is structured as follows.
In \cref{sec:symmetriesOfTheISR} we derive a connection between latent symmetries of $\ham$ and degeneracies in its eigenvalue spectrum.
\Cref{sec:localSymmetriesAndISR} relates latent symmetries of $\ham$ to local symmetries of $\ham$ in all matrix powers.
In \cref{sec:LatSymPresPerts} we develop the concept of complement multiplets, which allow one to perturb a Hamiltonian featuring a latent permutation symmetry without breaking this symmetry.
In \cref{sec:QMatrixFacts} we provide details on the generalized exchange symmetries.
\Cref{sec:bandstructure} shows the derivation of the Bloch Hamiltonians for the lattices of Figs. 1 and 2 of the main text.
\Cref{sec:AuxLemmata} contains auxiliary Lemmata used in proofs of Theorems in this Supplemental Material.

Throughout the following, $\lambdaDomain$ denotes the set of values of $\lambda$ for which the isospectral reduction $\ISR{S}{\ham}$ is defined.

\section{The relation between the symmetries of the isospectral reduction and degenerate eigenvalues of the Hamiltonian} \label{sec:symmetriesOfTheISR}
In the main part of this work, it was stated that non-abelian latent symmetries of a Hamiltonian necessarily induce degeneracies onto its eigenvalue spectrum.
A key in proving this statement lies in the application of representation theoretical tools to the isospectral reduction, which we shall do in the following
\begin{Theorem}{Symmetries of the isospectral reduction and degeneracies of $\ham$}{ISRSymmetries}
	Let $\ISR{S}{\ham}$ be the isospectral reduction of the Hamiltonian $\ham$ over a set of sites $S$, and let $G$ be a finite group with elements $\{g\}$ represented by matrices $\{\Gamma(g) \}$, and let $\ISR{S}{\ham}$ commute with all of them, i.e.,
$[\ISR{S}{\ham}, \Gamma(g)] = 0 \;\forall\; \lambda \in \lambdaDomain, r \in G $.

	Let $\Gamma$ be decomposed into $n$ pairwise non-equivalent irreducible representations $\widetilde{\Gamma}_{i}$ of $G$ with multiplicities $a_{i}\ne 0$ and with dimensions $d_{i}$, that is, there exists an invertible matrix $A$ such that
	\begin{equation}
	\Gamma'(g) = A \, \Gamma(g) \, A^{-1} = \bigoplus_{i=1}^{n} \widetilde{\Gamma}_{i}^{\oplus a_{i}}(g) \; \forall \; g \in G
	\end{equation}
	where $\oplus$ denotes the direct sum, and 
	$M^{\oplus k} = \underbrace{M \oplus \ldots{} \oplus M}_{k-\text{times}}$.
	Then for each $a_{i}$ the eigenvalue spectrum of $\ham$ contains \emph{at least} $a_{i}$ eigenvalues that are (individually) $d_{i}$-fold degenerate.
\end{Theorem}

\begin{proof}
	$\ISR{S}{\ham}$ represents a whole family of matrices parametric in $\lambda$. Each matrix in $\ISR{S}{\ham}$ commutes with the representation $\Gamma(g)$ of each group element $g$ of the finite symmetry group $G$.
	Thus, employing Schur's lemma, it is easy to prove that for each $\lambda \in \lambdaDomain$, $\ISR{S}{\ham}$ is block-diagonalized by the same similarity transformation
	\begin{equation} \label{eq:BlockDiagonalizationPreStep}
		\mathcal{R'}_{S}(\ham,\lambda) = A \, \ISR{S}{\ham} \, A^{-1} = \bigoplus_{i=1}^{n} B_{i}(\lambda)
	\end{equation}
	with $B_{i}(\lambda)$ being a $(a_{i}\, d_{i})$-dimensional matrix.
	Moreover, due to Schur's lemma, each $B_{i}(\lambda)$ can be further block-diagonalized by permuting its rows and columns, thereby yielding
	\begin{equation} \label{eq:BlockDiagonalization}
	\mathcal{R''}_{S}(\ham,\lambda) = P \, \mathcal{R'}_{S}(\ham,\lambda) \, P^{-1} = \bigoplus_{i=1}^{n} b_{i}^{\oplus d_{i}}(\lambda)
	\end{equation}
	with $P$ denoting the corresponding permutation matrix, and $b_{i}(\lambda)$ a matrix of dimension $a_{i}$.
	
	If we denote by $\mathbb{W}_{\pi}$ the set of rational functions $p(\lambda)/q(\lambda)$ with the numerator degree being less than or equal to the denominator degree, then every matrix element $(\mathcal{R}_{S})_{i,j} \in \mathbb{W}_{\pi}$ \cite{Bunimovich2014IsospectralTransformationsNewApproach}.
	Moreover, since $\mathbb{W}_{\pi}$ is closed under linear transformations, $(\mathcal{R''}_{S})_{i,j} \in \mathbb{W}_{\pi}$. This means that every $a_i$-dimensional block $b_{i}(\lambda)$ of $\mathcal{R''}_{S}$, when being solved for its non-linear eigenvalues via $\det\left(\mathcal{R''}_{S}(\ham,\lambda) - \lambda I \right) = 0$, features \emph{at least} $a_i$ solutions \cite{Bunimovich2014IsospectralTransformationsNewApproach}. 
	Since $\mathcal{R''}_{S}$ features $d_i$ such blocks and since every eigenvalue of $\ISR{S}{\ham}$ is also an eigenvalue of $\ham$ \cite{Bunimovich2014IsospectralTransformationsNewApproach}, it follows that for each irreducible representation $\widetilde{\Gamma}_{i}$ of dimension $d_{i}$ and multiplicity $a_i$ the Hamiltonian $\ham$ contains at least $a_{i}$ eigenvalues that are (individually) $d_{i}$-fold degenerate.
\end{proof}

\section{The connection between symmetries of the isospectral reduction and local symmetries of $\ham^k$} \label{sec:localSymmetriesAndISR}
One of the main results of this work is the relation between latent symmetries of a matrix $\ham$ and its local symmetries \emph{in every matrix power}.
This result was given without proof in the main part of this work, and is proven below.

\begin{Theorem}{}{commutationWithA}
	Let $\ISR{S}{\ham}$ denote the isospectral reduction of $\ham$ over some set of sites $S$, and let $A \in \mathbb{C}^{|S| \times |S|}$, where $|S|$ denotes the number of sites in $S$.
	Then
	\begin{equation} \label{eq:commutationWithHS,S}
	\comm{A}{\ISR{S}{\ham}} = 0 \;\forall\; \lambda \in \lambdaDomain \;\; \Leftrightarrow \;\; \comm{A}{\left(\ham^{k} \right)_{S,S}} = 0 \;\forall\;k \ge 0,
	\end{equation}
	where $(H^k)_{S,S}$ denotes the submatrix derived from $\ham^{k}$ by taking the rows and columns corresponding to the set $S$.
\end{Theorem}

\begin{proof}	
	``$\Rightarrow$'' can be shown by induction.
	The initial case $k=0$ is trivially fulfilled.
	The next step is to assume $\comm{A}{\left(\ham^{k} \right)_{S,S}} = 0$ holds for all $0 \le k \le k'$
	and then to show it holds also for $0 \le k \le k'+1$.
	By partitioning the matrix $H$ into blocks over $S$ and its complement $\bar{S}$ 
	the following identity can be shown to hold:
	\begin{align}  
	\left(\ham^{k} \right)_{S,S} = \left(\ham^{k-1} \right)_{S,S} \ham_{S,S} + \sum_{m=0}^{k-2} \left(\ham^{m} \right)_{S,S} \ham_{\ssb} \left(\ham_{\sbb} \right)^{k-2-m} \ham_{\sbs} \label{eq:HDecomposition}.
	\end{align}	
	Evaluating the commutator of $A$ with \cref{eq:HDecomposition} for $k=k'+1$ and applying the induction assumption along with \cref{eq:commutationWithAInEachPower} of \cref{lem:commutationWithA} we get $\comm{A}{\left(\ham^{k+1} \right)_{S,S}} = 0$, which completes the induction.
	
	``$\Leftarrow$'': First, we evaluate the commutator of $A$ with \cref{eq:HDecomposition} 
	and apply the assumption $\comm{A}{\left(\ham^{k} \right)_{S,S}} = 0 \;\forall\;k \ge 0$. 
	Next, again by induction we can show that 
	$\comm{A}{\ham_{\ssb} \left(\ham_{\sbb} \right)^{l} \ham_{\sbs}} = 0 \;\forall\; l \ge 0$.
	To prove that $\comm{A}{\ISR{S}{\ham}} = 0$ for all $\lambda \in \lambdaDomain$ we use the identity \cref{lemma3} for $\ISR{S}{\ham}$ from \cref{lem:ISRInTermsOfCharacteristicPolynomial}. Since $\comm{A}{\ham_{S,S}} = 0$ by assumption and $\comm{A}{\ham_{\ssb} \left(\ham_{\sbb} \right)^{n} \ham_{\sbs}} = 0 \;\forall\; n \ge 0$ by induction, we have that $\comm{A}{\ISR{S}{\ham}} = 0$ for all $\lambda \in \lambdaDomain$.
\end{proof}

\section{Modifications preserving latent permutation symmetries} \label{sec:LatSymPresPerts}
Given a Hamiltonian $\ham \in \mathbb{C}^{N\times N}$ with a latent permutation symmetry, it is often possible to modify $\ham$ while keeping this symmetry.
In particular, by analyzing the matrix powers of $\ham$, a large class of such latent-symmetry-preserving modifications can be found, as we derive in the following. 
We will start by defining what we call complement multiplets.
\begin{Definition}{Complement multiplet}{sbarMultiplets}
	Let $S$ be a set of sites of a hermitean Hamiltonian $\ham \in \mathbb{C}^{N \times N}$, that is, $S \subseteq \{1,\ldots{},N\}$, and $\sbar{}$ denote its complement (i.e., all other sites of $\ham$).
	A set $\mplet$ of sites of $\ham$ with $\mplet \subseteq \sbar$ forms a complement multiplet with respect to $S$ if
	\begin{equation} \label{eq:sbarMultipletDef}
	\sum_{m \in \mplet} \left(H \,  \overline{\ham}^{k} \right)_{s, m} = c_{k} \in \mathbb{C} \; \forall \; s \in S, k \ge 0 .
	\end{equation}
	where $\overline{\ham}$ is obtained from $\ham$ by setting the couplings between $S$ and $\sbar$ to zero.
\end{Definition}
Once a (subset of) complement multiplets have been identified, they can be used to modify the Hamiltonian without breaking the underlying latent symmetry, with the procedure and its proof detailed in the following
\begin{Theorem}{}{multipletKeepLatentSymmetry}
	Let $S$ be a set of sites of the hermitian Hamiltonian $\circAbove{\ham} \in \mathbb{C}^{N\times N}$.
	If one modifies $\circAbove{\ham} \rightarrow H\in \mathbb{C}^{(N+1)\times{}(N+1)}$ by adding a single site $c$ (with arbitrary on-site potential) and subsequently coupling each complement multiplet $\mplet_{j}$ of $\circAbove{\ham}$ to the site $c$ with the coupling $h_{j}$, i.e.,
	\begin{equation}
	\label{eq:couplings}
	\ham_{x,c} = \ham_{c,x}^{*} = \sum_{x \in \mplet_{j}} h_{j} ,
	\end{equation}
	with the star denoting complex conjugate,
	then the isospectral reduction changes as
	\begin{equation} 
	\label{eq:ISRChange}
	\mathcal{R}_{S}(H,\lambda) = \mathcal{R}_{S}(\circAbove{\ham},\lambda) + a(\lambda) J,
	\end{equation}
	with $a(\lambda)$ being a rational function in $\lambda$, $J \in \mathbb{R}^{|S| \times |S|}$ is a matrix of ones, and where $|S|$ denotes the number of sites in the set $S$.
	In particular, if $S$ is latently permutation symmetric in $\circAbove{\ham}$, it remains latently permutation symmetric in $\ham$. 
\end{Theorem}

\begin{proof}
	In order to show \cref{eq:ISRChange} holds, we evaluate the difference
	between $\ISR{S}{\ham}$ and $\ISR{S}{\hamCirc}$.
	To this end, we define $\hamTilde$ as the matrix obtained 
	from $\circAbove{\ham}$ by adding the site $c$ \emph{without connecting it}, 
	and by \cref{lem:blocksDoNotChangeRS},
	$\ISR{S}{\hamCirc} = \ISR{S}{\hamTilde}$.
	Further, we denote $\Sbarc= \sbar \cup c$.
	Next, for a sufficiently large $\lambda_{0}$ and $\lvert \lambda \rvert >\lambda_{0}> 0$, the matrix inverses occurring in $\ISR{S}{\ham}$ and $\ISR{S}{\hamTilde}$
	can be simultaneously formulated as convergent Neumann series.
	Finally, we note that $\ham_{\Ssb}=\hamTilde_{\Ssb}$ and $\ham_{S,c}=\hamTilde_{S,c}=0$.
	We arrive at:
	\begin{align}
		\ISR{S}{\ham} - \ISR{S}{\circAbove{\ham}}
		=& \sum_{k=1}^{\infty} \ham_{\Ssb} \left[\left( \ham_{\Sbbc} \right)^{k-1} - \left( \hamTilde_{\Sbbc} \right)^{k-1} \right] \ham_{\Sbs} t^{k} \\
		=& \sum_{k=1}^{\infty} \circAbove{\ham}_{\ssb}\left[ \left( \ham_{\sbb} \right)^{k-1} - \left( \hamTilde_{\sbb} \right)^{k-1} \right]\circAbove{\ham}_{\sbs} t^{k} \label{eq:isr_diff}
	\end{align}
	where $t=1/\lambda$.
	In the following, we abbreviate the terms in square brackets of \cref{eq:isr_diff} by the matrix 
	$\Delta^{(k-1)}_{\sbb}$.
	We further denote by $\overline{\ham}$ and $\circAbove{\overline{\ham}}$ the matrices obtained from $\ham$ and $\circAbove{\ham}$, respectively, by decoupling the set of sites $S$ from the remaining sites. 
	
	Using a graph-theoretical interpretation (see Ref. \cite{Morfonios2020aCospectralityPreservingGraphModifications}) the $(i,j)$-th matrix element of $\Delta^{(k)}_{\sbb}$ can be expressed in terms of \emph{walks} in a graph $\mathbb{G}(\overline{\ham})$ of length $k$ starting at site $\overline{s}_{i}$, that is, the $i$-th element of $\sbar$, and ending at site $\overline{s}_{j}$ while necessarily visiting the new site $c$ at least once.
	This yields
	\begin{align} \label{eq:summands}
	\Delta^{(k)}_{\sbb} =& \sum_{2+l+m+n=k} \left[\left( \sum_{x \in \cup_i \mplet_{i}} \left(\circAbove{\overline{\ham}}^{l} \right)_{\sbar{},x} \ham_{x,c} \right) \left(\overline{\ham}^{m} \right)_{c,c} \left( \sum_{x \in \cup_i \mplet_{i}} \ham_{c,x} \left(\circAbove{\overline{\ham}}^{n} \right)_{x,\sbar{}}  \right) \right]
	\end{align}
	
	Next, we replace $\ham_{x,c}$ by \cref{eq:couplings} and multiply \cref{eq:summands} with  $\circAbove{\ham}_{\ssb}$ on the left
	and its transpose on the right (compare with \cref{eq:isr_diff}):
	\begin{align}
	\label{eq:multiply_to_see_mplet}
	& \sum_{2+l+m+n=k} \left[\left(\sum_{i} \sum_{x \in \mplet_{i}} \circAbove{\ham}_{\ssb} \left(\circAbove{\overline{\ham}}^{l} \right)_{\sbar{},x} h_{i} \right) \left( \overline{\ham}^{m} \right)_{c,c} \left(\sum_{i} \sum_{x \in \mplet_{i}} h_{i}^{*} \left(\circAbove{\overline{\ham}}^{n} \right)_{x,\sbar{}} \circAbove{\ham}_{\sbs}  \right) \right]
	\end{align}
	
	Using \cref{eq:sbarMultipletDef}, we recognize that $\sum_{x \in \mplet_{i}} \circAbove{\ham}_{\ssb} \left(\circAbove{\overline{\ham}}^{l} \right)_{\sbar{},x}=c_{l}^{(i)} \vec{1}_S$ with an-all constant $|S|$-dimensional vector $\vec{1}_S$.
	This means that the matrix in \cref{eq:multiply_to_see_mplet} is proportional to all-one matrix $J$ with a constant pre-factor $a^{(k)}=\sum_{2+l+m+n=k} \left[\left(\sum_{i} h_{i} c_{l}^{(i)} \right) \left( \overline{\ham}^{m} \right)_{c,c} \left(\sum_{i} h_{i} c_{n}^{(i)}  \right)^{*} \right]$.
	Inserting this result in \cref{eq:isr_diff}, we obtain that $a(\lambda)$ from \cref{eq:ISRChange} equals $\sum_{k=1}^{\infty} a^{(k-1)} \left(\frac{1}{\lambda}\right)^k$ for $\lvert \lambda \rvert > \lambda_{0}$.

	Finally, we note that the left-hand side of \cref{eq:isr_diff} is a rational function of $\lambda$.
	Since it equals $a(\lambda) \, J\;\forall\; \lvert \lambda \rvert >\lambda_0$, by the identity theorem for polynomials it must be equal to $a(\lambda) \, J$ also for $\lvert \lambda \rvert \le \lambda_{0}, \; \lambda \; \in \lambdaDomain$.
	Now, since $J$ commutes with any permutation matrix, each latent symmetry of $\circAbove{\ham}$ is preserved in $\ham$.
\end{proof}

\section{The connection between (latent) $C_{n>2}$ symmetries and degeneracies for real-symmetric Hamiltonians} \label{sec:QMatrixFacts}
In the following, we will present more details on the concept of generalized exchange symmetry (GES).
We will then use these to finally prove \cref{thm:latentCNSymmetriesInducesLatentDihedralSymmetry}, which states that, for real-valued Hamiltonians, a more than twofold rotational symmetry (that is, a symmetry $C_{n>2}$) necessarily leads to a non-abelian latent $D_{n}$ symmetry.

As explained in the main part of this work, each generalized exchange symmetry (GES) is given by a symmetric orthogonal matrix $\qMat{u}{v}$ which permutes two sites $u,v$ while acting as an orthogonal transformation on the others.
The $\qMat{u}{v}$ were introduced in Ref. \cite{Godsil2017A1StronglyCospectralVertices}, where it has been shown that, for real Hamiltonians, $\qMat{u}{v}$ exists if and only if $(\ham^{k})_{u,u} = (\ham^{k})_{v,v}$ for all $k$.
In this case, the eigenvalue spectra of $\ham \setminus u$ and $\ham \setminus v$ coincide, and the two sites $u$ and $v$ are said to be cospectral \cite{Godsil2017A1StronglyCospectralVertices}.

Similar to Ref. \cite{Chan2020AQFundamentalsFractionalRevivalGraphs}, we will now explicitly construct $\qMat{u}{v}$ by means of projectors.
To this end, we first choose the eigenstates according to the following
\begin{Lemma}{\hypersetup{citecolor=white} Lemma 2.5 of Ref. \cite{Eisenberg2019DM3422821PrettyGoodQuantumState}}{eisenberg}
	Let $\ham$ be a real symmetric matrix, with $u$ and $v$ cospectral.
	Then the eigenstates $\{\ket{\phi}\}$ of $\ham$ are (or, in the case of degenerate eigenvalues, can be chosen) as follows. 
	For each eigenvalue $\lambda$ there is at most one eigenstate $\ket{\phi}$ with even local parity on $u$ and $v$, i.\,e., $\braket{u|\phi_{i}^{(+)}} = \braket{v|\phi_{i}^{(+)}} \ne 0$, and at most one eigenstate $\phi$ with odd local parity on $u$ and $v$, i.\,e., $\braket{u|\phi_{i}^{(-)}} = - \braket{v|\phi_{i}^{(-)}} \ne 0$.
	Here, $\ket{x}$ denotes a vector which possesses the value one at site $x$ and zeros on all other sites.
	All remaining eigenstates for $\lambda$ fulfill $\braket{u|\phi_{i}^{(0)}} = \braket{v|\phi_{i}^{(0)}} = 0$.
	The even (odd) parity eigenstate can be found by projecting the vector $\ket{u} \pm \ket{v}$ onto the eigenspace associated with $\lambda$.
\end{Lemma}
With this choice of the eigenstate basis, we state the following
\begin{Theorem}{}{GES_theo}
	Let the orthonormal eigenstates $\{\ket{\phi}\}$ of $\ham$ be chosen according to \cref{lem:eisenberg}, and define the projectors
	\begin{equation}
		\label{eq:plusspace}
		P_{+}^{(u,v)} = \sum_{i} \ket{\phi_{i}^{(+)}}\bra{\phi_{i}^{(+)}}, \quad P_{-}^{(u,v)} = \sum_{i} \ket{\phi_{i}^{(-)}}\bra{\phi_{i}^{(-)}}, \; P_{0}^{(u,v)} = \sum_{i} \ket{\phi_{i}^{(0)}}\bra{\phi_{i}^{(0)}}
	\end{equation}
	Then $\qMat{u}{v} = P_{+}^{(u,v)} + P_{0}^{(u,v)} - P_{-}^{(u,v)}$
	fulfills
	\begin{equation}
		(\qMat{u}{v})^{-1} = (\qMat{u}{v})^{T} = \qMat{u}{v}, \;\; \qMat{u}{v} \ket{u} = \ket{v} .
	\end{equation}
\end{Theorem}
\begin{proof}
The property $(\qMat{u}{v})^{-1} = \qMat{u}{v}$ follows simply from the fact that the projection matrices \cref{eq:plusspace} are idempotent.
$(\qMat{u}{v})^{T} = \qMat{u}{v}$ follows from the fact that one can choose the eigenvectors of $\ham$ to be real-valued, so that the projector onto the eigenspace associated to any eigenvalue is real, thereby rendering also the projection matrices \cref{eq:plusspace} real and therefore also symmetric.

In order to prove that $\qMat{u}{v} \ket{u} = \ket{v}$, we use \cref{lem:eisenberg} and the orthonormality of eigenstates $\ket{\phi_i}$
to get $\braket{v|\qMat{u}{v}|u} = (\qMat{u}{v})_{u,v} =\sum_i \braket{u|\phi_{i}}\braket{\phi_{i}|u}=1$.
Additionally, since $\sum_{i} (\qMat{u}{v}_{u,i})^2 = \sum_{i} (\qMat{u}{v}_{v,i})^2 = 1$ due to the orthogonality of $Q^{(u,v)}$, it follows that $(\qMat{u}{v})_{u,i} = \delta_{i,v}$ and $(\qMat{u}{v})_{v,i} = \delta_{i,u}$.
\end{proof}

With these prerequisites and a good understanding of the concept of GES, we are now finally able to prove the connection between $C_{n}$ rotational symmetries of a real Hamiltonian and the necessary emergence of $D_{n}$ latent permutation symmetries, explicated in the following
\begin{Theorem}{}{latentCNSymmetriesInducesLatentDihedralSymmetry}
	Let $\ham \in \mathbb{R}^{N \times N}$ be a real symmetric Hamiltonian that features a latent or non-latent $C_{n>2}$ permutation symmetry.
	Then
	\begin{itemize}
		\item $\ham$ necessarily also features a latent $D_{n}$ permutation symmetry and features at least $\floor{\frac{n-1}{2}}$ pairs of doubly degenerate eigenvalues, where $\floor{x}$ rounds $x$ down to the nearest integer.
		\item There exist two GESs of $\ham$ which do not commute with each other.
	\end{itemize}
\end{Theorem}
\begin{proof} 	
	\begin{itemize}
		\item If $\ham$ features a (latent or non-latent) $C_{n>2}$ permutation symmetry, then there is at least one set $S$ of $n$ sites and a $n\times n$ permutation matrix $P$ fulfilling
		\begin{equation} \label{eq:conditionsOnP}
		P^{k} \ne I \;\forall\; 1 \le k < n,\;\;P^n = I,\;\;\; \comm{\ISR{S}{\ham}}{P} = 0 \;\forall\; \lambda \in \lambdaDomain
		\end{equation}
		where $I$ is the identity matrix.
		Together with the symmetry and real-valuedness of $\ham$, this property implies that the rows and column of $\ISR{S}{\ham}$ can be permuted such that it is a real symmetric circulant matrix. It is known that such matrices commute with the permutation matrix corresponding to the operation that performs a flip about the anti-diagonal. Together with the cyclic permutations of order $n$, this operation generates the dihedral group $D_{n}$, and $\ham$ thus features a latent $D_{n}$ permutation symmetry.
		
		Next, we note that it is known that the eigenstates 
		of real symmetric circulant matrices are independent of their entries
		\cite{Tee2005RLIMS8123EigenvectorsBlockCirculantAlternating},
		here in particular independent of $\lambda$.
		Using them to diagonalize $\ISR{S}{\ham}$ one obtains a diagonal $n\times n$ matrix with entries $ f_{j}(\lambda) \in \mathbb{W}_{\pi} \;, j=1,\ldots{},n$, that is, rational functions $p_{j}(\lambda)/q_{j}(\lambda)$
		with the numerator degree being less 
		than or equal to the degree of the denominator.
		The eigenvalues of $\ISR{S}{\ham}$ are thus given by the sum of the multisets denoting the solutions 
		to $f_{j}(\lambda) - \lambda = 0$.
		Similar to the proof of \cref{thm:ISRSymmetries}, it can be shown that each of these equations has at least one solution.
		Furthermore, $f_{j}(\lambda) = f_{n-j}(\lambda)$, because $\ISR{S}{\ham}$ is not only circulant but also real-symmetric.
		Finally, since every eigenvalue of $\ISR{S}{\ham}$ is contained in the spectrum of $\ham$ \cite{Bunimovich2014IsospectralTransformationsNewApproach}, we conclude that $\ham$ has at least $\floor{\frac{n-1}{2}}$ pairs of doubly degenerate eigenvalues.
		
		\item From the above and from \cref{thm:commutationWithA}
		we have $\comm{\left(\ham^{k} \right)_{S,S}}{P} = 0$ for all $k\ge 0$, 
		i.e. $\left( \ham^{k} \right)_{S,S}$ is a real symmetric circulant matrix.
		In each power $k$, the diagonal elements $\left(\ham^{k} \right)_{ii} = \left(\ham^{k} \right)_{jj} \; \forall \; i,j$ are pairwise equal, meaning that each pair of sites in $S$ is cospectral.
		By Ref. \cite{Godsil2017A1StronglyCospectralVertices}, for each such pair there is a GES $\qMat{i}{j}$ which commutes with $\ham$ and \cref{thm:GES_theo} applies.
		
		For the sake of simple notation let us now assume that the sites of $\ham$ are labeled (if this is not the case, one can easily renumber the sites accordingly) such that $s_i \rightarrow i$, with $P$ permuting the sites in $S = \{s_{1},\ldots{},s_{|S|}\}$ such that the site $i<n$ is mapped onto the site $i+1$, and $n$ onto $1$.
		The fact that $(\ham^{k})_{S,S}$ is circulant symmetric then implies $(\ham^{k})_{1,2} = (\ham^{k})_{2,3} \;\forall\; k$.
		In the terminology of Ref. \cite{Morfonios2020aCospectralityPreservingGraphModifications}, site $2$ is a walk-singlet w.r.t.\ the cospectral sites $1$ and $3$.
		Thus, by Theorem 4 from Ref. \cite{Morfonios2020aCospectralityPreservingGraphModifications}, 
		the eigenstates of $\ham$ [chosen according to \cref{lem:eisenberg} 
		for the cospectral pair $1$ and $3$] with negative parity 
		on cospectral sites vanish on site $2$.
		By combining the projector definition of $\qMat{1}{3}$ with the completeness relation of eigenstates one can show $(\qMat{1}{3})_{2,2} = 1$.
		Furthermore, since $\qMat{1}{3}$ is orthogonal, the matrix elements
		$(\qMat{1}{3})_{2,j} = \delta_{2,j}$.
		On the other hand, site $3$ is not necessarily a singlet 
		for the  cospectral pair $1$ and $2$.
		Specifically, we have
		\begin{equation}
		\qMat{1}{3} = \left(\begin{array}{@{}c|c@{}}
		\begin{matrix}
		0 & 0 & 1 \\
		0 & 1 & 0 \\
		1 & 0 & 0
		\end{matrix}
		& \bigThing{0} \\
		\hline 
		\bigThing{0} &
		\begin{matrix}
		\bigThing{A}
		\end{matrix}
		\end{array}\right),
		\;\;
		\qMat{1}{2} = \left(\begin{array}{@{}c|c|c@{}}
		\begin{matrix}
		0 & 1  \\
		1 & 0  \\
		\end{matrix}
		& \bigThing{0} & \bigThing{0} \\
		\hline
		\bigThing{0} & a & b\\
		\hline
		\bigThing{0} & b^{T} & \bigThing{B}
		\end{array}\right)
		\end{equation}
		where $A,B \in \mathbb{R}^{(N-3)\times{}(N-3)}$ and $b \in \mathbb{R}^{1\times{}(N-3)}$.
		Since the upper left $3 \times 3$ block of the commutator of the above two matrices does not vanish, these two matrices do not commute.
	\end{itemize}
\end{proof}

\section{Band-structure calculations} \label{sec:bandstructure}

To derive the Bloch-Hamiltonian $\ham_{B}(\mathbf{k}) = \ham_{B}(k_{x},k_{y})$, we follow the convention of Eq. (2.75) of Ref. \cite{Vanderbilt2018BerryPhasesElectronicStructure}.
To this end, let $\ket{m_{\mathbf{R}}}$ denote the state which is completely localized on site $m$ of the unit cell located at position $\mathbf{R}$.
For our two-dimensional lattices, the vectors $\mathbf{R} = A \vec{a}_{1} + B \vec{a}_{2}$, where $A,B$ are integers and $\vec{a}_{1,2}$ are the two primitive vectors describing the lattice.
The Bloch-Hamiltonian can then be written as
\begin{equation} \label{eq:BlochHamiltonian}
	\left(\ham_{B}(\mathbf{k}) \right)_{nm} = \sum_{\mathbf{R}} e^{i \mathbf{k} \cdot \mathbf{R}} \braket{m_{\mathbf{0}}|\ham_{L}|m_{\mathbf{R}}}
\end{equation}
where $\ham_{L}$ denotes the Hamiltonian of the infinite lattice.

\subsection{Bloch-Hamiltonian for the modified Kagome lattice}
With $\vec{a}_{1} = (1,0)^{T}$, $\vec{a}_{2} = (\frac{1}{2}, \frac{\sqrt{3}}{2})^{T}$, we obtain the Bloch-Hamiltonian of Fig. 1 (c) of the main text as
\begin{equation}
	\left(
	\begin{array}{cccccc}
	0 & h_{3} & h_{3} & h_{2} & 0 &
	h_{1} \\
	h_{3} & 0 & h_{3} & h_{1} &
	h_{2} & 0 \\
	h_{3} & h_{3} & 0 & 0 & h_{1} &
	h_{2} \\
	h_{2} & h_{1} & 0 & 0 & h_{4}
	e^{i \left(\frac{\sqrt{3}
			k_{y}}{2}-\frac{k_{x}}{2}\right)}
	& h_{4} e^{-i k_{x}} \\
	0 & h_{2} & h_{1} & h_{4} e^{-i
		\left(\frac{\sqrt{3}
			k_{y}}{2}-\frac{k_{x}}{2}\right)}
	& 0 & h_{4} e^{-i
		\left(\frac{k_{x}}{2}+\frac{\sqrt{3}
			k_{y}}{2}\right)} \\
	h_{1} & 0 & h_{2} & h_{4} e^{i
		k_{x}} & h_{4} e^{i
		\left(\frac{k_{x}}{2}+\frac{\sqrt{3}
			k_{y}}{2}\right)} & 0 \\
	\end{array}
	\right) \, .
\end{equation}

At $k_{x} = k_{y} = 0$, the isospectral reduction of the Bloch-Hamiltonian over the sites $S=\{1,2,3 \}$ then has the structure
\begin{equation}
	\begin{pmatrix}
	a(\lambda) & b(\lambda) & b(\lambda) \\
	b(\lambda) & a(\lambda) & b(\lambda) \\
	b(\lambda) & b(\lambda) & a(\lambda)
	\end{pmatrix} .
\end{equation}
Thus, the Bloch-Hamiltonian features a latent $D_{3}$ permutation symmetry at $\mathbf{k} = 0$.

\subsection{Construction of the Bloch-Hamiltonian belonging to Fig. 2 (b)}
Let us here briefly discuss how one could build a lattice from the Hamiltonian depicted in Fig. 2 (a) of the main text.
The basic idea is to interpret $\ham$ as the Bloch-Hamiltonian $\ham_{B}$ of an extended lattice, evaluated at the $\Gamma$-point, that is, $\ham = \ham_{B}(\mathbf{k} = 0)$.
Out of the many possible ways to achieve this, we here first remove the three curved couplings---the ones between sites $(6,11)$, $(5,10)$, and $(4,7)$---from $\ham$, and use the resulting system $\ham_{UC}$ as the unit cell of a lattice, as depicted in Fig. 2 (b).
The corresponding Bloch-Hamiltonian $\ham_{B}(\mathbf{k}) = \ham_{UC} + \ham_{IC}(\mathbf{k})$ is the sum of the unit cell Hamiltonian $\ham_{UC}$ and a $\mathbf{k}$-dependent inter-cell coupling $\ham_{IC}(\mathbf{k})$.
The matrix elements of $\ham_{IC}(\mathbf{k})$ are obtained by taking an arbitrary reference unit cell, $A$.
For each site $i$ in $A$ that is connected with coupling $h_{ij}$ to a site $j$ in an adjacent unit cell $B$, $\left(\ham_{IC}(\mathbf{k}) \right)_{ij} = h_{ij}\, e^{i \mathbf{k} \cdot \mathbf{R}_{AB}}$ with $\mathbf{R}_{AB}$ denoting the vector pointing from $A$ to $B$, as one can show by means of \cref{eq:BlochHamiltonian}.
At $\mathbf{k} = 0$, all complex exponentials in $\ham_{IC}(\mathbf{k})$ become unity, and one can design $\ham_{B}(\mathbf{k=0}) = \ham$ by suitably connecting the initial unit cell to its neighbors.
For the above choice and by chosing $\vec{a}_{1} = (0,1)^{T}$, $\vec{a}_{2} = (\frac{5}{3} \cos(\pi/6),\frac{1}{2})^{T}$, we then obtain the Bloch-Hamiltonian of Fig. 2 (b) of the main text as
\begin{equation}
	\begin{pmatrix}
	\mathbf{0}_{6\times 6} & \mathbf{C} \\
	\mathbf{C}^{\dagger} & \mathbf{0}_{5\times 5}
	\end{pmatrix}
\end{equation}
with $\mathbf{0}_{n\times n}$ denoting the $n\times n$ matrix of zeros, and
\begin{equation}
	\mathbf{C} = \begin{pmatrix}
	h & h & 0 & 0 & h'' \\
	h & 0 & h & h' & 0 \\
	2 h & h & h & 0 & 0 \\
	h e^{i k_{y}} & 2 h & 2 h & 0 & 0 \\
	h & 0 & h' e^{i\left(\frac{k_{y}}{2}-\frac{5 k_{x}}{2 \sqrt{3}}\right)} & 0 & 0 \\
	0 & 0 & h & 0 & h'' e^{i(\frac{5 k_{x}}{2 \sqrt{3}} + \frac{k_{y}}{2})}
	\end{pmatrix} .
\end{equation}
For $k_{x} = k_{y} = 0$, this Hamiltonian equals the one depicted in Fig. 2 (a) of the main text, as intended.

\section{Auxiliary Lemmata} \label{sec:AuxLemmata}
\begin{Lemma}{}{blocksDoNotChangeRS}
	Let $S$ be a set of sites of $\circAbove{\ham}$. If one extends
	\begin{equation}
	\circAbove{\ham} \rightarrow H = \begin{pmatrix}
	\circAbove{\ham} & 0 \\
	0 & H'
	\end{pmatrix}
	\end{equation}
	then the isospectral reduction over $S$ remains unchanged, i.e.,
	$\mathcal{R}_{S}(H,\lambda) = \mathcal{R}_{S}(\circAbove{\ham},\lambda)$ .
\end{Lemma}
\begin{proof}
	Follows straightforwardly from the fact that 
	$\left(\ham_{\sbar,\sbar}- \lambda I\right)^{-1} = \begin{pmatrix}
	\left(\ham_{\sbar_{1},\sbar_{1}} - \lambda I\right)^{-1} & 0 \\
	0 && \left(\ham_{\sbar_{2},\sbar_{2}} - \lambda I\right)^{-1}
	\end{pmatrix}$
	is block diagonal and $\ham_{S,\sbar_{2}}=0$ with sites $\sbar_{1}$ 
	from $ \circAbove{\ham} $, $\sbar_{2}$ from $H'$ and $\sbar_{1} \cup \sbar_{2} = \sbar$.
\end{proof}

\begin{Lemma}{}{commutationWithA}
	Let the isospectral reduction
	\begin{equation} 
	\ISR{S}{\ham} = \ham_{S,S} - \ham_{\ssb} \left( \ham_{\sbb} - \lambda I \right)^{-1} \ham_{\sbs}
	\end{equation}
	over some set $S$ of sites.
	If $\ISR{S}{\ham}$ commutes with a $|S| \times |S|$ matrix $A$, that is,
	$\comm{A}{\ISR{S}{\ham}} = 0 \;\forall\; \lambda \in \lambdaDomain$, then
\begin{flalign}
	& \comm{A}{\ham_{S,S}} = 0 \label{eq:HSSCommutes}\\
	& \comm{A}{\ham_{\ssb} \left(\ham_{\sbb} - \lambda I \right)^{-1} \ham_{\sbs}} 
	= 0 \;\forall\; \lambda \in \lambdaDomain. \label{eq:commutationWithA}\\
	& \comm{A}{\ham_{\ssb} \left(\ham_{\sbb}\right)^{k} \ham_{\sbs}} 
	= 0 \;\forall\; k \ge 0 \label{eq:commutationWithAInEachPower}.
\end{flalign}

\end{Lemma}
\begin{proof}
	First, \cref{eq:HSSCommutes} follows from $\comm{A}{\ISR{S}{\ham}} = 0$ by evaluating the limit $\lambda \rightarrow \infty$.
	Second, \cref{eq:commutationWithA} follows from \cref{eq:HSSCommutes} and $\comm{A}{\ISR{S}{\ham}} = 0$.
	Last, for a sufficiently large $\lambda_{0}$ and $\lvert \lambda \rvert >\lambda_0 > 0$, the left-hand side of \cref{eq:commutationWithA} can be formulated as a convergent power series in $x=\frac{1}{\lambda}$:
	\begin{equation}
	\frac{1}{\lambda_{0}}\sum_{k=1}^{\infty} c_k x^{k} = 0 \;\forall\; x: 0< \lvert x \rvert < \frac{1}{\lambda_0}
	\end{equation}
	where $c_{k} = \comm{A}{\ham_{\ssb} \left(\frac{\ham_{\sbb}}{\lambda_{0}}\right)^{k-1} \ham_{\sbs}}$.
	By the identity theorem for power series it follows that all $c_{k} = 0$, 
	thus proving \cref{eq:commutationWithAInEachPower}.
\end{proof}

\begin{Lemma}{}{ISRInTermsOfCharacteristicPolynomial}
	Let
	$\ISR{S}{\ham} = \ham_{S,S} - \ham_{\ssb} \left(\ham_{\sbb} - \lambda I \right)^{-1} \ham_{\sbs}$
	be the isospectral reduction of $\ham$ over a site set $S$.
	Then
	\begin{equation}
	\ISR{S}{\ham} = \ham_{S,S} + \sum_{k=1}^{|\sbar|} \frac{c_{k}}{c_{0}} \sum_{n=0}^{k-1} \mybinom{k-1}{n}\left(- \lambda \right)^{k-1-n} \ham_{\ssb} \left(\ham_{\sbb} \right)^{n} \ham_{\sbs} \;\forall\;\lambda \in \lambdaDomain
	\label{lemma3}
	\end{equation}
	where $c_{i} = c_{i}(\lambda)$ are the coefficients of the characteristic polynomial of $\ham_{\sbb} - \lambda I$.
\end{Lemma}
\begin{proof}
	For $\lambda \in \lambdaDomain$ the matrix $\ham_{\sbb} - \lambda I:= M$ is invertible, with characteristic polynomial
	$p_{M}(\lambda,x) = \sum_{k=0}^{|\sbar|} c_{k}(\lambda) x^{k}$ and $c_0 \neq 0$.
	From the Cayley-Hamilton theorem one obtains the identity relation
	$M^{-1} = - \sum_{k=1}^{|\sbar|} \frac{c_{k}}{c_{0}} M^{k-1}$.
	Inserting this relation into the definition of isospectral reduction and 
	applying the binomial theorem for $\left(\ham_{\sbb} - \lambda I\right)^{k-1}$ we arrive at \cref{lemma3}.
\end{proof}